\documentclass[%
 reprint,
 amsmath,amssymb,
 aps,
]{revtex4-1}

\usepackage{graphicx}
\usepackage{dcolumn}
\usepackage{bm}

\begin{document}


\title{Spatial squeezing in bright twin beams generated with four-wave mixing: constraints on characterization with an EMCCD camera}

\author{Ashok Kumar$^{1,2}$ and  A. M. Marino$^1${\footnote{marino@ou.edu}}}
\affiliation{$^1$Homer L. Dodge Department of Physics and Astronomy, The University of Oklahoma, Norman, Oklahoma 73019, USA}
\affiliation{$^2$Department of Physics, Indian Institute of Space Science and Technology, Thiruvananthapuram, Kerala 695547, India}
\begin{abstract}
The observation of spatial quantum noise reduction, or spatial squeezing, with a large number of photons can lead to a significant advantage in quantum imaging and quantum metrology due to the scaling of the signal-to-noise ratio with the number of photons. Here we present a systematic study of the limiting factors that play a role on the measurement of spatial squeezing with an electron-multiplying charge coupled device (EMCCD) camera in the limit of bright quantum states of light generated with a four-wave mixing process in an atomic vapor cell.  We detect a total number of photons per beam of the order 10$^8$ in 1 $\mu$s pulses, which corresponds to a photon flux per beam of the order of 10$^{14}$ photons per second. We then investigate the role of different parameters, such as cell temperature, pump power, laser detunings, scattered pump background noise, and timing sequences  for the image acquisition with the EMCCD camera, on the level of spatial squeezing. We identify critical parameters to obtain an optimum squeezing level and demonstrate that for bright beams it is essential to acquire images at a rate fast enough to overcome the effect of classical technical noise.
\end{abstract}
\maketitle

\section{INTRODUCTION}

Spatial quantum correlations that lead to spatial squeezing can play an important role in quantum metrology and quantum imaging as they make it possible to perform measurements below the shot noise limit~\cite{Serg1, RMP, Serg2, Treps1, Kolobov, Walborn, Brida, Serg3}. Furthermore, spatial quantum correlations are useful in quantum information processing  as they provide  high-dimensional entangled photons that can encode information in transverse spatial modes~\cite{Padgett,Fleischer}. As a result, the study of spatial quantum correlations has become an active area of research~\cite{Padgett, Lantz, Genovese, Lee, Wasilewski, Boyer, Fleischer}. While most of the work on spatial squeezing has focused on photon-pairs generated through spontaneous parametric down conversion (SPDC)~\cite{shot1, shot2, shot3}, we have recently shown that it is possible to obtain spatial squeezing in the regime of a large number of photons with bright twin beams generated with a four-wave mixing (FWM) process~\cite{Kumar}. As opposed to measurements done in the photon-pair regime, the ability to measure spatial quantum correlations with bright beams makes it possible to obtain sub-shot noise imaging in a single shot with a significant level of noise reduction. Moreover, the ability to generate quantum states of light with a macroscopic number of photons can lead to further enhancements in sensing applications due to the $\sqrt{N}$ scaling of the signal-to-noise ratio, where $N$ is the number of photons.

In this paper, we present a systematic study of the limiting factors that play a role in the use of an EMCCD camera to observe spatial squeezing in bright twin beams of light generated with FMW in a double-$\Lambda$ configuration in a hot rubidium vapor cell. The use of this process as a source allows us to generate quantum correlated photons without the need of an optical cavity~\cite{Lett1, Glorieux}, which makes it possible to preserve the multi-spatial-mode nature of the source. It also provides the added advantage of a controllable mean photon number over a wide range while keeping the level of quantum correlations fixed. These properties make the quantum correlated twin beams generated with this source a viable option for quantum metrology, quantum imaging, and quantum information processing applications~\cite{Lett2, Jing, Pooser, Dowran}.

The manuscript is organized as follows. In section II, we discuss the details of the experiment, data acquisition, and data analysis. We present the experimental results in section III, which is divided into three subsections. In subsection III.A, we focus on the effect of  different parameters of the FWM on the measured level of spatial squeezing. In particular, for a given pump-probe pulse timing sequence, we study the dependence of the spatial squeezing on the cell temperature, pump power, and laser detuning. Once we have determined the optimum operational parameters of the source, we keep them fixed for the results presented in the subsequent sections. In subsection III.B, we discuss the contribution of scattered background photons, which add noise to the measurement, on the measured level of spatial squeezing. We show that for large number of photocounts in the analysis region, the scattered pump photons do not degrade the spatial squeezing significantly, while they do have a significant impact at a lower number of  photocounts. In subsection III.C, we study the dependence of the spatial squeezing on timing effects. In particular, we consider pump-probe pulse timing sequences, which affect transient effects in the FWM, and camera acquisition rates, which affect detection, for the optimized cell temperature, pump power, and one photon detuning. We present a simple model to emulate the effect of excess noise on the measured level of spatial squeezing. Finally, we conclude in section IV.

\section{EXPERIMENT}

A schematic of the experimental setup is shown in Fig.~1. We generate bright twin beams of light with a FWM process in a hot $^{85}$Rb vapor cell at a temperature of 110$^{\circ}$C. A strong pump beam, with a power of 2.15~W and a $1/e^2$ diameter of 4.4~mm, intersects at an angle of 0.4~degrees inside the cell with a weak probe beam, with a power in the $\mu$W level and a $1/e^2$ diameter of 0.4~mm.  Both the pump and the probe are derived from the same Ti-Sapphire laser, which is frequency locked to a reference vapor cell. The probe beam is generated by taking a small portion of the pump and using an acousto-optic modulator (AOM) to downshift the frequency by 3.04~GHz with respect to the pump frequency, such that the pump and probe are on two-photon resonance with the hyperfine ground levels of ${^{85}}$Rb, as shown in the inset of Fig.~1. To precisely control the one-photon detuning $\Delta$ of the pump laser, we use a saturated absorption spectroscopy setup to lock the laser frequency at a given detuning from the $F = 2$ to $F' = 3$ transition in the ${^{85}}$Rb D$_1$ line.

\begin{figure}[hbt]
\centering
\includegraphics[width=\columnwidth]{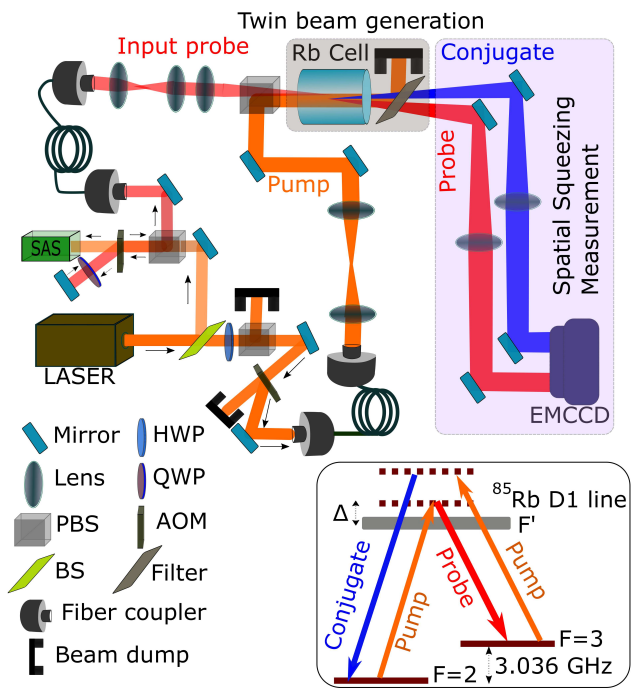}
\caption{Experimental setup for the characterization of spatial squeezing in the far field. HWP: Half wave plate; QWP: Quarter wave plate, AOM: Acousto-optic modulator, PBS: Polarizing beam splitter, BS: Beam splitter, SAS: Saturated absorption spectroscopy, EMCCD: Electron-multiplying charge coupled device.}
\label{fig:Exp}
\end{figure}

As a result of the FWM process in a double-$\Lambda$ configuration (see inset in Fig.~1), the input probe beam is amplified and a new beam, the conjugate, is generated. The simultaneous generation of probe and conjugate photons (twin beams) by the FWM process leads to temporal quantum correlation between them while phase matching, or the conservation of momentum, leads to spatial quantum correlations or spatial squeezing. In our current experiment we limit our study of these spatial quantum correlations to the far field. In order to measure the spatial squeezing in this regime, the probe and conjugate beams are sent through two separate lenses, each of focal length $f=500$~mm.  The lenses are used in an $f$-to-$f$ configuration to obtain the Fourier transform of the fields at the cell center on the EMCCD camera~(ProEM-HS: 512BX3). To minimize the amount of pump light that reaches the EMCCD, the pump beam is filtered after the cell with a polarization filter.

The generation of bright twin beams requires an input probe beam to seed the FWM process. This introduces a DC-gaussian profile and classical excess noise on the generated probe and conjugate beams \cite{Serg4}. Given that the spatial quantum correlations manifest themselves in the relative spatial intensity fluctuations between the fields \cite{Knight}, it is necessary to extract these small spatial fluctuations from the large DC-gaussian background, and to eliminate the classical excess noise introduced by the seed as much as possible in order to observe the spatial squeezing. To do so, we acquire two consecutive sets of probe and conjugate images in a short time interval. The subtraction of these two sets of images allows us to obtain the spatial intensity fluctuation for each beam as well as to cancel out any temporal classical technical noise up to a frequency of the order of the inverse time between the two sets of images. Thus, it is necessary to acquire the sets of images as fast as possible.  We do this through the use of the kinetic mode of the EMCCD camera, which allows us to capture multiple frames containing  probe and conjugate images in fast succession in a single acquisition from the camera. The EMCCD camera that we use consists of an active area of 512$\times$512~pixels and a buffer region of 512$\times$512~pixels for storage. We divide the total sensor area into six frames of equal size such that each frame is 170$\times$512~pixels. For the analysis, we use two of the frames that are stored in the buffer region, which helps to minimize contributions from background and scattered light. Given that the EMCCD camera has a vertical charge shift rate of 300~ns/row, the time between frame can be made as small 51~$\mu$s.

\begin{figure}[hbt]
\centering
\includegraphics{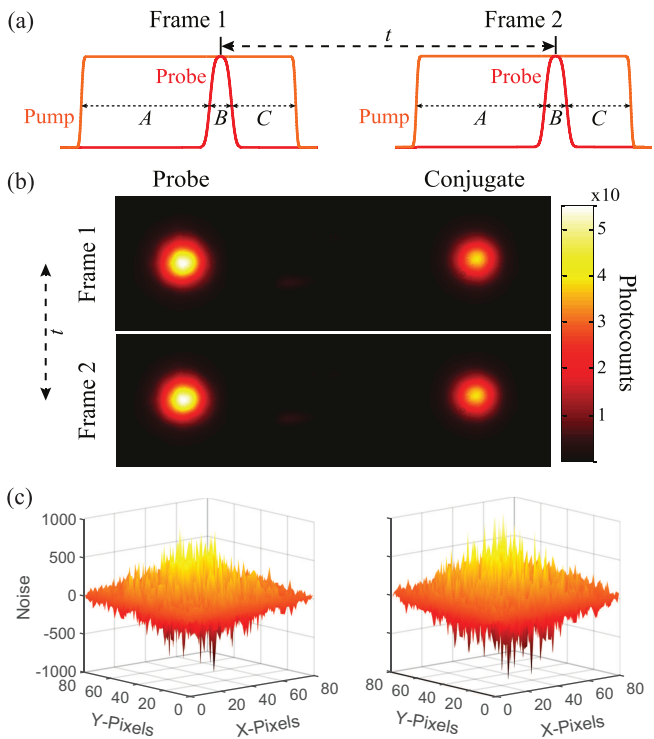}
\caption{Image acquisition of probe and conjugate pulses with the EMCCD camera. (a)~Pulse sequence for the input probe and pump for the acquisition of probe and conjugate images in two consecutive frames of the EMCCD camera. (b)~Images of probe and conjugate pulses in consecutive frames. Each frame is $170\times512$ pixels, with a pixel size of $16\times16~\mu$m. The time interval $t$ between the two frames can be controlled (see text for the details). (c)~Probe and conjugate spatial noise (spatial photocount fluctuations) obtained after the subtraction of two consecutive frames for an analysis region of 80$\times$80 pixels.}
\label{fig:TimeSequence}
\end{figure}

For the image acquisition, the input probe and pump beams are pulsed with AOMs to implement the pulsing sequence shown in Fig.~\ref{fig:TimeSequence}(a). The timing of the pulse sequence is synchronized with the acquisition of the frames in the EMCCD by triggering the AOMs with the camera such that an image of single probe and conjugate pulses are acquired in each of the 6 frames. For a typical timing sequence in our experiment, the pulse width of the pump ($A+B+C$) is kept at 10~$\mu$s, while the probe pulse width ($B$) is set to 1~$\mu$s. The probe pulse is turned on after a delay ($A$) of 6~$\mu$s with respect to the pump pulse to avoid any transient effects in the FWM, which will be discussed in detail in next sections.

We record images of the twin beams for various settings of different parameters that affect the FWM process, such as cell temperature, one-photon detuning, pump power, and pump-probe pulse timing sequence; and the detection with the EMCCD, such as vertical charge shift rate of the EMCCD camera. For each measurement, we capture 100 image sequences, with each image sequence containing a total of 6 frames. To study the effect of the scattered pump, we record images of the scattered pump background noise by turning the input seed off and following the same procedure used to acquire the probe and conjugate images. To obtain an accurate measure of the scattered pump noise, we acquire a background noise image after each probe-conjugate image acquisition.

Figure~\ref{fig:TimeSequence}(b) shows two consecutive frames, each with a size of 170$\times$512 pixels and with an image of a probe and conjugate pulse. We subtract these images to obtain images of the spatial intensity fluctuations of the probe and the conjugate, as shown in Fig.~\ref{fig:TimeSequence}(c). To do so, a section of 120$\times$120 pixels around the maximum intensity region of the probe and conjugate in each image are cropped and aligned in each frame with an image registration algorithm. We then further crop a section of 80$\times$80 pixels area around the maximum intensity region in each aligned probe and conjugate images for the final noise analysis. Due to thermal effects of the AOM there is a small drop in the power of the seed probe for the second frame with respect to first one, and thus of the total photocounts from one frame to the other.  As a result, a scaling factor of $\sim$~1.003, which is obtained by taking the ratio between the total probe photocounts in the analysis regions in the two frames, is applied to the second frame to rescale its intensity before the two frames are subtracted. We have verified that this scaling factor is consistent for every consecutive frame used in the analysis. The scaling factor effectively balances the different input probe seed powers and makes it possible to completely eliminate the DC-gaussian profiles of the probe and conjugate images when performing the subtraction between the frames.  Without proper balancing, there would be a spatially dependent offset in the spatial noise shown in Fig.~\ref{fig:TimeSequence}(c), which would bias the measured levels of noise. Once the images of the intensity spatial fluctuations of the probe and conjugate are obtained by subtracting the two frames, see Fig.~\ref{fig:TimeSequence}(c), they can be used to calculate the spatial squeezing by subtracting the corresponding pixels and performing spatial statistics. Due to the phase matching condition, the spatially correlated regions between the probe and conjugate are located diametrically opposite to each other, therefore it is necessary to rotate the image of the intensity fluctuations of the conjugate by 180~degrees with respect to the one of the probe before subtracting the two.

We define the noise ratio ($\sigma$) to characterize the spatial squeezing in the bright twin beams as follows
\begin{equation}\label{NRF}
    \sigma \equiv \frac{\langle \delta^2[(N_{p1}-N_{p2})-(N_{c1}-N_{c2})] \rangle}{\langle N_{p1}+N_{c1}+N_{p2}+N_{c2} \rangle},
\end{equation}
\noindent where $N_{p1}$, $N_{p2}$, $N_{c1}$, and $N_{c2}$ are the matrices representing the photocounts per pixel for probe and conjugate images in two consecutive frames and the  statistics are performed over the pixels of the images. In this expression, $(N_{p1}-N_{p2})$ and $(N_{c1}-N_{c2})$ give the images of the spatial intensity fluctuations of the probe and conjugate, respectively, as shown in Fig.~\ref{fig:TimeSequence}(c), such that the numerator represents the spatial variance of the intensity difference noise between the probe and conjugate. The denominator gives the mean photocounts for the probe plus conjugate pulses used for the analysis and represents the shot noise.  For coherent state pulses $\sigma=1$, which corresponds to the shot noise limit, while for thermal light or other classical states $\sigma>1$. Spatially quantum correlated beams, like the twin beams generated in our experiment, will result in $\sigma<1$ when the contribution from the quantum correlations to $\sigma$ can be made to dominate over the contributions from classical excess noise. Thus, the noise ratio ($\sigma$) can directly quantify the spatial squeezing, with a smaller $\sigma$ corresponding to a larger degree of spatial squeezing or spatial quantum correlations.

It is important to note that as a result of the non-planner (gaussian) pump profile used for the FWM process, there is a spatial spread, known as the coherence area, between the correlated regions of the probe and conjugate spatial intensity fluctuations. The coherence area provides a minimum spatial scale for the spatial correlations and can be directly measured through the cross-correlation between the spatial intensity fluctuations of the probe and the conjugate~\cite{KumarCoh}. For our current optical system, we have measured the full width at half maximum of the coherence area to be $\sim$10$\times$10 pixels. In order to observe a significant degree of spatial squeezing, the size of a pixel in the detection area needs to be larger than the size of the coherence area, as partial detection of a coherence area by different pixels is equivalent to attenuation that degrades the spatial squeezing~\cite{Kumar}. In particular, in the limit where the pixel size is smaller than the coherence area, the effect is equivalent to spatially cutting a single spatial mode beam ~\cite{MarinoCoh}. Therefore, in order to properly characterize the level of spatial squeezing with our current optical system, we need to perform the spatial statistics by grouping or binning pixels to define ``superpixels''. As we bin the pixels, we cut less and less of the coherence area, which leads to the spatial squeezing increasing and then saturating as the size of a super pixel exceeds the size of the coherence area. To illustrate this fact, we analyze the spatial squeezing, or noise ratio defined in Eq.~(1), as a function of the number of pixels that are binned in each superpixel.

There is, however, a tradeoff between binning a large number of pixels to obtain an accurate measure of the level of spatial squeezing and having enough superpixels in the analysis region to perform the statistics required for the noise ratio.  It is thus necessary to have as many coherence areas as possible for a given analysis region~\cite{Gatti}. This means that the size of the coherence area should be made as small as possible relative to the size of the probe and conjugate in the far field. Recently we have shown that the size of the coherence area in the twin beams can be controlled by changing the pump beam size~\cite{MarinoCoh}. In particular, an increase in the pump size reduces the size of coherence area.  Therefore, we use a pump beam  with a $1/e^2$ diameter of 4.4~mm, with its size determined by a compromise between a reduction in the size of the coherence area and the available pump power (2.15~W). The use of such a large pump beam leads to an increase of the pump power required to get a large enough FWM gain ($\sim 4.0$ in this experiment), which in turn increases the scattered pump background noise.

\section{RESULTS AND DISCUSSION}

In this section, we discuss the results obtained from scanning over the different FWM and detection parameters that play a role in the level of spatial squeezing. For all the results that we present, we calculate the noise ratio defined in Eq.~(1) as a function of the number of binned pixels.  To perform the binning, we group a square region of $n \times n$~pixels in the computer to  form the superpixels. To keep the total number of superpixels an integer, we used a slightly different analysis region for each binning.

Unless noted, the total number of photocounts in an analysis region of 80$\times$80~pixels is kept at about 10$^{8}$. Given that images are captured with an input probe pulse with a 1~$\mu$s duration, the total photocounts per second in each beam is of the order of 10$^{14}$. For the seeded FWM process, in the absence of losses and competing effects, every input photon generates $G-1$ correlated photon pairs, where $G$ is the gain of the process. For our experimental configuration $G=4$, which means that 3/4 of all the measured photons are quantum correlated. This translates to a flux of quantum correlated photons of $\sim7.5\times10^{13}$ per second in each beam. It is important to note that the total number of photocounts is limited by saturation of the EMCCD camera and not by the FWM source.

\subsection{Role of cell temperature, pump power, and one-photon detuning}

\begin{figure*}[t]
\centering
\includegraphics{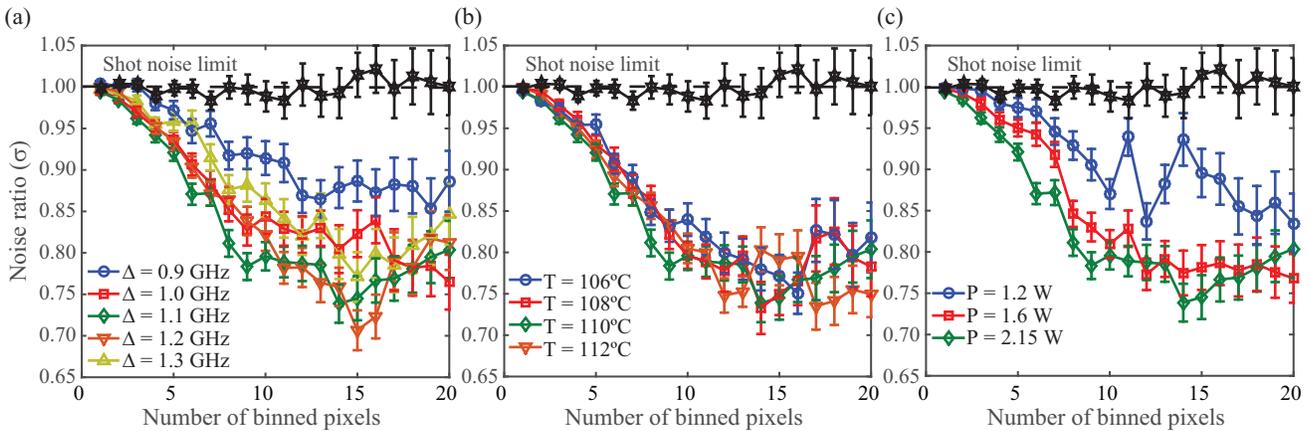}
\caption{Measured noise ratio as a function of the number of binned pixels for different values of (a) one-photon detuning ($\Delta$) at a cell temperature ($T$) of 110$^{\circ}$C and pump power ($P$) of 2.15~W, (b) cell temperature for a one-photon detuning of 1.1~GHz and pump power of 2.15~W, and (c) pump power for a one-photon detuning of 1.1~GHz and a cell temperature of 110$^{\circ}$C. The black trace shows the measured noise ratio with two coherent laser beams and corresponds to the shot noise limit. The error bars in all the plots represent the standard deviation of the mean noise ratio over the acquired 100 shots.}
\label{fig:DetTempPow}
\end{figure*}

We first study the effect of source (FWM) parameters on the observed level of spatial squeezing. More specifically, we consider the effect of temperature ($T$), pump power ($P$), and one-photon detuning ($\Delta$). We vary the cell temperature from 106$^{\circ}$C to 112$^{\circ}$C, and for each temperature, we study the effect of one-photon detuning and pump power on the spatial squeezing. We find the maximum spatial squeezing for a cell temperature of 110$^{\circ}$C, one-photon detuning of 1.1~GHz, and pump power of 2.15~W. In Fig.~3, we plot the measured noise ratio for different combinations of one-photon detuning, cell temperature, and pump power. More specifically, Fig.~\ref{fig:DetTempPow}(a) shows  the noise ratios for values of $\Delta$ from 0.9~GHz to 1.3~GHz at a fixed cell temperature of 110$^{\circ}$C and pump power of 2.15~W. As can be seen, the maximum noise reduction is obtained for $\Delta$=1.1~GHz, detuning at which the noise ratio decreases more sharply at lower binning. The dependence of the noise reduction on $\Delta$ can be attributed to the fact that for lower values of $\Delta$, the probe frequency comes closer to  atomic resonances, which leads to probe absorption and as a result to a degradation of the level of squeezing. For larger values of $\Delta$, the probe frequency moves away from atomic resonance, which leads to a reduction of the FWM gain and as a result of the level of squeezing. Therefore, one has to operate the FWM at an optimum value of $\Delta$ where we have a balance between the minimum probe absorption and the maximum FWM gain. Figure~\ref{fig:DetTempPow}(b) shows how the noise ratio changes for cell temperatures ranging from 106$^{\circ}$C to 112$^{\circ}$C for an optimized one-photon detuning of 1.1~GHz and a pump power of 2.15~W.  As can be seen in  Fig.~\ref{fig:DetTempPow}(b), within the range we considered, the temperature of the cell does not have a significant effect on the noise reduction, nevertheless the optimum temperature appears to be 110$^{\circ}$C given slightly faster reduction in the noise ratio for low binning.

Finally, we scan over a range of pump powers to study the dependence of the spatial squeezing on this parameter.  We plot the noise ratio for pump powers values of 1.2~W, 1.6~W, and 2.15~W at an optimized cell temperature of 110$^{\circ}$C and $\Delta$ of 1.1~GHz in Fig.~\ref{fig:DetTempPow}(c). As the pump power is decreased, the FWM gain decreases and as a result the level of spatial squeezing decreases. Thus, when comparing all the results shown in Fig.~3, we can see that for our pump size, angle between pump and probe, and two-photon detuning, an optimum spatial squeezing is obtained at a cell temperature of 110$^{\circ}$C, pump power of 2.15~W, and one-photon detuning of 1.1~GHz. The data presented in the subsequent sections is taken with these optimized parameters.

The overall behavior of the noise ratio with binning seen in Fig.~3 can be understood, as explained before, as an effective loss due to spatially cutting coherence areas that occurs when the size of a superpixel is smaller or of the order of the coherence area.  This behavior is consistent with the model we previously developed to explain the noise of the twin beams as they are clipped \cite{KumarCoh, MarinoCoh}. As can be seen from Fig.~3, once the size of the superpixel exceeds the size of the coherence area, the noise ratio saturates at $\sim$0.7. This level of spatial noise reduction (spatial squeezing) is limited by the level of intensity difference squeezing in the temporal domain, which for our experiment is of $\sim$0.3.  The reduction in spatial squeezing is due in part to the lower quantum efficiency of the EMCCD camera, which is of 70$\%$. This leads to an expected level of squeezing of ~0.5.  Additionally, the discrepancy between the expected and measured levels of spatial squeezing can be understood by taking into account the fact that when images are taken with the EMCCD we are effectively integrating over a region of the squeezing spectrum over which the intensity difference squeezing is reduced from its maximum value.

To validate our analysis and show that values of the noise ratio below one represent noise levels below the shot noise limit, or squeezing, we also perform the noise analysis with a classical coherent laser beam. To do so, we split a laser beam into two with a beam splitter and perform the same analysis and pulsing sequence as for the probe and conjugate beams. From these measurements we obtain a noise ratio ($\sigma$), defined in Eq.~(1), of $\sim$1 (see black trace in Fig.~\ref{fig:DetTempPow}), as expected for a classical coherent state.  This level corresponds to the shot noise limit and any noise ratio below this level is a sufficient condition for the presence of quantum correlations between the probe and conjugate. This result also shows how the fast acquisition of subsequent frames used for the analysis is able to cancel any classical technical noise present in the light.

\subsection{Role of scattered pump noise}

The measured level of spatial squeezing can degrade significantly in the presence of scattered background photons that reach the EMCCD as the noise from these photons adds in quadrature with the reduced noise level from the twin beam. In our experiment, the frequencies of the generated twin beams are close to that of the pump beam ($\sim 3$~GHz away from each of the twin beams) as shown in the inset of Fig.~1.  Therefore, it is not possible to eliminate the scattered pump photons via an interference filter and the use of a cavity would modify or filter out the spatial correlations. Even though we can separate the orthogonally polarized pump beam and twin beams with a polarization filter, due to the large power required for the pump there are still residual scattered pump photons that make it to the EMCCD. These scattered pump photons are one of the main sources of excess noise in our experiment. Therefore, it is important to study the role of this source of noise on the measured level of spatial squeezing.

We study the effect of the scattered pump photons by comparing the noise ratio calculated with and without taking the background noise into account. For each of the 100 probe-conjugate sequences, we take a corresponding background sequence and use the same analysis region as the one used for the corresponding probe-conjugate image for the data analysis. We then take the background noise into account by defining the following background subtracted noise ratio

\begin{widetext}
 \begin{equation}
   \sigma_{B} \equiv \frac{\langle \delta^2[(N_{p1}-N_{p2})-(N_{c1}-N_{c2})]\rangle-\langle \delta^2[(N_{pb1}-N_{pb2})-(N_{cb1}-N_{cb2})]\rangle}{\langle N_{p1}+N_{c1}+N_{p2}+N_{c2} \rangle-\langle N_{pb1}+N_{bc1}+N_{bp2}+N_{bc2} \rangle},
    \end{equation}
 \end{widetext}
where ($N_{pb1}$, $N_{cb1}$) and ($N_{pb2}$, $N_{cb2}$) are the matrices representing the photocounts per pixel for the background noise images in the two consecutive frames used for the analysis.

\begin{figure}[hbt]
\centering
\includegraphics{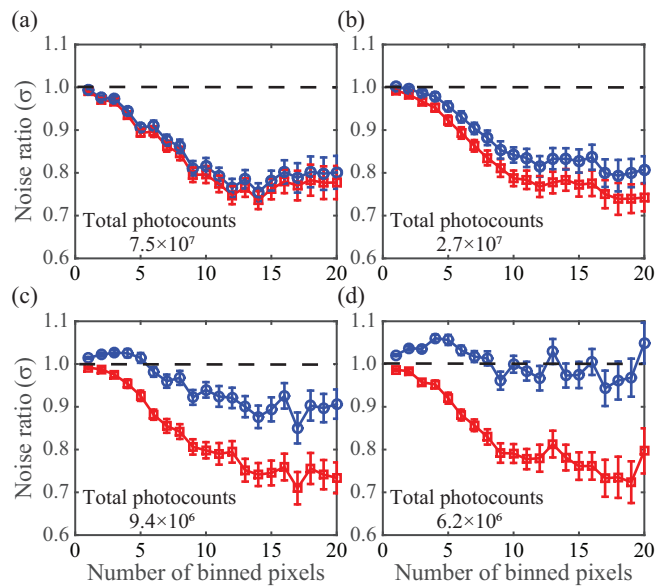}
\caption{Effect of scattered pump photons on the measured level of spatial squeezing. We calculate the noise ratio with (red traces, $\square$) and without (blue traces, $\circ$) background noise subtraction for a total of (a) $7.5\times10^7$, (b) $2.7\times10^7$, (c) $9.4\times10^6$, and (d) $6.2\times10^6$ photocounts in the analysis region.}
\label{fig:BGNoise}
\end{figure}

To study the impact of the scattered pump photons, we perform experiments where we change the total number of photocounts by changing the power of the input probe beam while keeping the power of the pump fixed.  Figure~\ref{fig:BGNoise} shows the noise ratio with and without background noise correction for these measurements. The total number of probe photocounts in Figs.~\ref{fig:BGNoise}(a) to (d) in our analysis region with a 1~$\mu$s long probe pulse are $7.5\times10^7$, $2.7\times10^7$, $9.4\times10^6$, and $6.2\times10^6$, respectively, with the maximum number of photocounts limited by saturation of the EMCCD.  For all these measurements the total background scattered photocounts in our analysis region are $\sim4\times10^5$. It is clear from Fig.~\ref{fig:BGNoise} that the contribution from the background noise becomes significant as the number of probe photocounts is reduced. At lower probe photocounts, the background noise not only degrades the spatial squeezing but can also completely dominate over it. However, it is possible to recover the spatial squeezing by subtracting the background noise through the use of Eq.~(2).  As can be seen in Fig.~\ref{fig:BGNoise}(d), for the lowest total photocount level we considered ($6.2\times10^6$), the level of spatial squeezing after background subtraction is the same as the one measured when the background noise does not play a significant role, see Fig.~\ref{fig:BGNoise}(a).

\subsection{Role of pump-probe timing sequence and camera acquisition rate}

As discussed in the experimental section, during the image acquisition, we use the pulse sequence shown in Fig.~\ref{fig:TimeSequence}(a) for the pump and the probe pulses. Here we study the role of the timing of the pump and probe pulse sequence on the observed level of spatial squeezing. We first focus on the role of transient effects in the FWM process by measuring the level of spatial squeezing for different delay times between the start of the pump and probe pulses; that is, for different value of $A$ in Fig.~\ref{fig:TimeSequence}(a) with fixed pump pulse duration ($A+B+C$) of 10~$\mu$s and probe pulse duration ($B$) of 1~$\mu$s. Given that for the ideal case of no loss, the noise ratio or squeezing depends on the gain $G$ of FWM and scales as $1/(2G-1)$,
we first directly look at the gain of the FWM process as a function of the probe-pump delay $A$. When we turn on the pump and the probe pulses simultaneously, there is no FWM gain given that it takes some time for the pump to prepare the atoms for the FWM. As the delay is increased, the FWM gain starts to increase after a delay of $\sim1~\mu$s and finally saturates at a value of $\sim4$ for our optimum experimental conditions  at a delay of $\sim6~\mu$s.

\begin{figure}[hbt]
\centering
\includegraphics{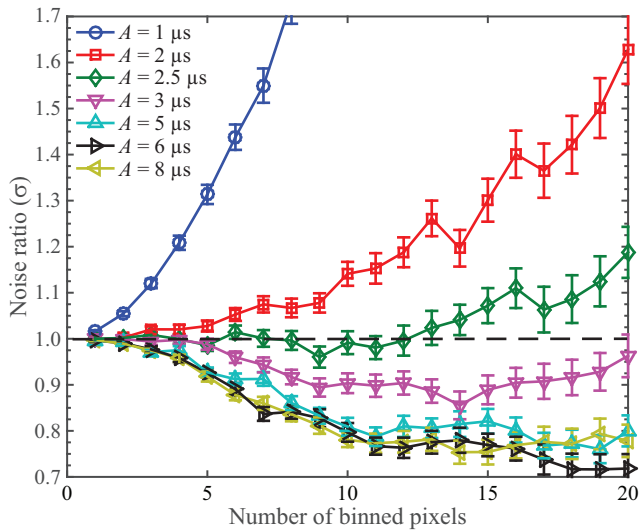}
\caption{Impact of FWM transient effects on the level of spatial squeezing. Measured noise ratio levels for different delays between the pump and probe pulses; that is, time $A$ in Fig.~\ref{fig:TimeSequence}(a). For these measurements the duration of the pump and probe pulses are kept constant at 10~$\mu$s and 1~$\mu$s, respectively.}
\label{fig:PumpProbePulseTimings}
\end{figure}

In order to see how the FWM transient effects, along with the resulting saturation of the gain, impact the spatial squeezing, we calculate the noise ratio for twin beam images acquired for different probe-pump delay times $A$, as shown in Fig.~\ref{fig:PumpProbePulseTimings}. It can be seen from this figure that for delays of the order of a few $\mu$s or less, for which the FWM gain is low, the spatial intensity fluctuations in the twin beams are dominated by the transient effects of the FWM process and present no squeezing. However, as we keep increasing the delay, we recover the spatial squeezing ($\sigma$<1). The noise ratio decreasing with increasing delay and saturates after a delay of about 5~$\mu$s to 6~$\mu$s, consistent with the increase and saturation of the FWM gain.  We have  verified that this behavior is independent of the pump pulse duration ($A+B+C$) and the probe pulse duration ($B$) and only depends on the delay between the pulses ($A$).
Thus, in order to avoid transient effects from the FWM process in our experiments, we turn on the probe pulse with a delay of 6~$\mu$s with respect to the pump pulse.

We also study the role of the acquisition rate of the images with the EMCCD camera on the noise reduction. To observe spatial squeezing with bright twin beams, it is essential to ``filter'' the DC gaussian profile introduced by the input seed beam, as the quantum correlations exist between the relative spatial intensity fluctuations of the twin beams. This can be accomplished by subtracting two images in consecutive frames acquired in rapid succession, which makes it possible to extract the spatial intensity fluctuations required to calculate the spatial squeezing. This technique also offers the advantage of canceling any low frequency spatial and temporal technical noise.  The ability to filter out the bright spatial DC gaussian profile is analogous to the technique routinely used in the time domain with a spectrum analyzer. In particular, in the temporal domain the spectrum analyzer makes it possible to filter out the bright DC portion of the beams as well as the low frequency technical noise since the quantum correlations or quantum noise reduction can be characterized at an analysis frequency different than DC \cite{Bachor}.

\begin{figure}[hbt]
\centering
\includegraphics{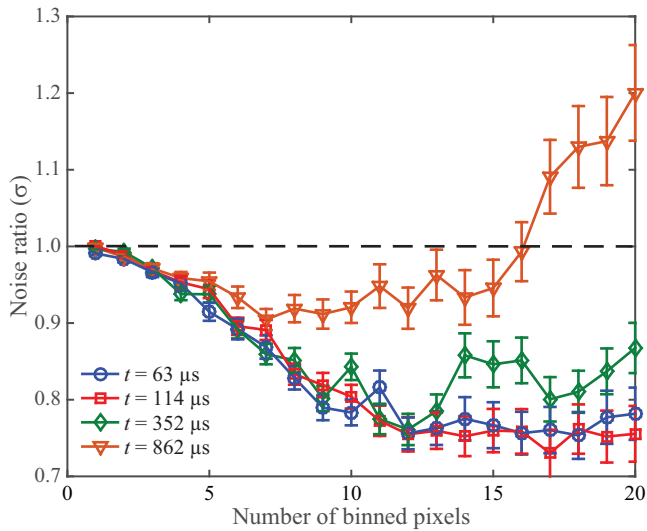}
\caption{Measured noise ratio as a function of binning for different vertical charge transfer rates of the EMCCD camera. A change in transfer rate leads to a change in the time between consecutive frames, $t$ in Fig.~\ref{fig:TimeSequence}. The noise ratio degrades with longer time between frames due to classical technical noise.}
\label{fig:ShiftRate}
\end{figure}

To effectively cancel the classical technical noise, it is necessary to acquire the two images to be subtracted in a time interval shorter than the inverse of the frequency of the noise we are trying to cancel. Thus, by keeping the time between frames as small as possible, we can subtract classical technical noise up to larger frequencies. We acquire two successive images of the probe and conjugate in a short time frame through the use of the kinetic mode feature of the EMCCD camera. Given that the EMCCD camera can transfer charge at a rate of 300~ns/row, 600~ns/row, 2000~ns/row, or 5000~ns/row, we can control the time interval between two consecutive frames. In addition to the time between frames, during our experiments we expose each frame of size 170$\times$512~pixels for a time of 12~$\mu$s, with a start and stop of the exposure 1~$\mu$s before and after the pump pulse. The combination of these two times allows us to take two images with a time between them ($t$ in Fig.~\ref{fig:TimeSequence}) of 63~$\mu$s, 114~$\mu$s, 352~$\mu$s, or 862~$\mu$s.  To study the effect of the technical noise on the spatial squeezing, we calculate the noise ratio for these four time intervals $t$ between images, see Fig.~\ref{fig:ShiftRate}. It can be seen from this figure that as we increase $t$ beyond a certain value, the spatial squeezing degrades and is even lost at higher binnings. Longer time intervals between probe and conjugate images can be achieved through the use of two non-consecutive  frames out of the six that are taken in each sequence. We have observed that when the time $t$ is larger than about 1~ms, no spatial squeezing is measured for any binning. This is due to the fact that with the larger time $t$, the cut off frequency for technical noise that can be canceled is lowered.  This leads to the excess technical noise dominating over the quantum spatial intensity fluctuations of the probe and conjugate, and thus making it impossible to see the quantum correlations through squeezing measurements.

In order to understand the source of the degradation of the spatial squeezing at longer time intervals $t$ and its dependence on binning, we develop a simple model for the noise. We generate two noise matrices of 80$\times$80~pixels, corresponding to the size of analysis region of the probe and conjugate, with random number between $-r$ and $r$ and zero mean. We then add these random noise matrices to the probe and conjugate spatial intensity fluctuation matrices obtained after subtraction of the two frames. For each probe-conjugate image sequence, we generate and add a different pair of random noise matrices. We take the probe and conjugate spatial intensity fluctuations obtained for the fastest  possible acquisition rate of the EMCCD camera ($t=63~\mu$s) as the baseline matrices we add the random noise to. Finally, we follow the same analysis as described above to calculate the noise ratio defined in Eq.~(\ref{NRF}). The addition of the noise matrices to the baseline probe and conjugate noise levels effectively models the technical noise that remains after subtraction of the two consecutive frames.

\begin{figure}[hbt]
\centering
\includegraphics{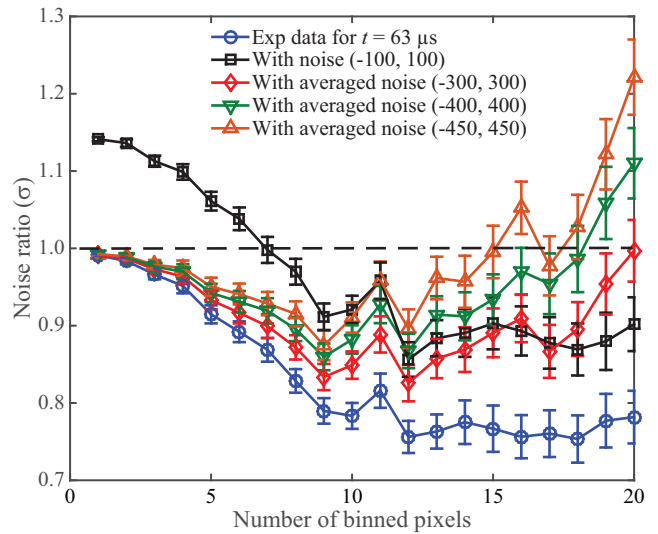}
\caption{Emulation of the role of classical technical noise on the noise ratio. The blue trace ($\circ$) serves as a baseline and is given by the experimental data acquired with the shortest possible time interval between the two images ($t=63~\mu$s). Random noise with values between the numbers in parenthesis are added to the baseline to model classical technical noise. The black trace ($\square$) gives the noise ratio when random noise is added at the pixel level to the baseline.  The peak-to-peak values of the noise (numbers in parenthesis) correspond to 7\% of the peak-to-peak spatial intensity fluctuations in the baseline. The red ($\diamond$), green ($\triangledown$), and orange ($\triangle$) traces give the noise ratios when a spatial running average is performed on the noise matrix before adding it to the baseline with peak-to-peak values of 21\%, 28\% and 32\% of the spatial intensity fluctuations of the baseline, respectively.}
\label{fig:ShiftRateSim}
\end{figure}

To model the effect of high spatial frequency classical noise between frames or spatial inhomogeneities in the quantum efficiency of the EMCCD, we directly add the noise matrices to the intensity fluctuation matrices to add random spatial noise at the pixel level. In this case the noise ratio shifts up as shown in Fig.~\ref{fig:ShiftRateSim} (black trace, $\square$ symbol) by a constant amount, which depends on the value of $r$, for all binnings with respect to the baseline or original noise ratio in Fig.~\ref{fig:ShiftRateSim} (blue trace, $\circ$ symbol).  Given that this result is inconsistent with the experimental results shown in Fig.~\ref{fig:ShiftRate}, where for longer time intervals $t$ the spatial squeezing degrades more at higher binnings than at lower binnings, we can conclude that noise either due to high frequency spatial fluctuations from shot to shot or inhomogeneities in the quantum efficiency of the pixels of the EMCCD camera do not play a significant role in our measurements. We have also independently verified that the pixel-to-pixel quantum efficiency inhomogeneities in our EMCCD camera do not introduce a significant amount of spatial noise in our measurements through the use of the flatfield correction technique given in Ref. \cite{Jiang}.

Another possible source of noise is temporal technical noise. To model this source of noise, we convolve the random noise matrices with a matrix composed of elements of equal value, normalized such that the sum of all its elements is equal to one, and of size equal to the full width at half maximum of the probe beam (i.e. $\sim$ 40$\times$40~pixels). The convolution effectively introduces classical spatial correlations to the noise matrix and makes it possible to model technical noise that is fairly uniform over the whole beam. After the convolution we add the noise matrices to the baseline probe and conjugate spatial intensity fluctuations as described above. We then calculate the noise ratios with different amplitudes for the added noise (values of $r$), as shown in Fig.~\ref{fig:ShiftRateSim} (traces with $\diamond$, $\triangledown$ and $\triangle$ symbols). In this model, a larger noise amplitude is equivalent to having a larger amount of noise not canceled by the two frame subtraction. Thus a larger noise amplitude is used to model a longer time $t$ between images. As can be seen from Fig.~\ref{fig:ShiftRateSim}, the noise ratios obtained with convolved spatially random noise follow a similar trend to the one seen in Fig.~\ref{fig:ShiftRate} as the binning and noise amplitude (or effectively time $t$) are increased. This results point to the high frequency temporal noise that is not canceled by the two frame subtraction technique as the source of noise that leads to the degradation of the spatial squeezing for longer times and binning of a larger number of pixels.

For the noise analysis presented here, it is also important to discuss the number of temporal and spatial modes involved in the analysis region. Regarding the spatial modes, with the current configuration of our experiment the number of spatial modes (as estimated from the extent of the coherence area) in Fig.~3 varies roughly from 0.01 to 4. It is important to note that the exact number of modes per super-pixel of the camera depends on the optical system used to perform the Fourier transform of the center of the cell to obtain the far field distribution on the EMCCD.  A more significant figure of merit is the number of spatial modes present in the whole analysis region of 80$\times$80 pixels, as we perform the spatial statistics over these modes. For our current configuration this is of the order of 64 spatial modes. As the bandwidth of the FWM process is of the order of 10 to 20 MHz, this leads to a typical number of temporal modes (probe pulse width/coherence time) within the 1~$\mu$s probe pulse width of the order of 10 to 20. If we consider the number of temporal modes and spatial modes detected in the analysis region, 64$\times$(10 to 20), we get a total number of modes detected per image of the probe and conjugate of the order of 1000. This means that each of the modes we are detecting has on average $10^5$ photons in the 1~$\mu$s capture time.

\section{CONCLUSION}

We  present a systematic analysis of the limiting factors for the use of an EMCCD camera to measure spatial squeezing in bright twin beams of light generated with FWM in an atomic vapor. We show the dependence of the spatial squeezing on various experimental parameters such as temperature of the Rb vapor cell, pump power, laser detunings from atomic resonances, number of probe photocounts, and pump-probe pulse timing sequences; as well as on detection parameters, such as camera acquisition rates. For our experimental setup, we find the optimum FWM parameters to obtain spatial squeezing to be a cell temperature of $\sim110^{o}$C, pump power of $\sim2.15$~W, and one-photon detuning of $\sim1.1$~GHz.  We also show that the role of background scattered pump photons becomes significant at a low number of probe photocounts, where the spatial squeezing can be completely suppressed by this source of noise, and that the spatial squeezing can be recovered by properly subtracting the background noise due to the scattered pump photons.

For the optimized FWM parameters (cell temperature, pump power, and laser detuning), we present a study of the effect of different pump and probe pulse timing sequences on the measured level of  spatial squeezing. We observe that an optimum noise reduction can be obtained when the probe pulse is delayed at least 6~$\mu$s with respect to the pump pulse in order to avoid transient effects in the FWM process. We have also studied the effect of different EMCCD acquisition rates on the spatial squeezing. We show that the spatial squeezing degrades as the image acquisition rate is decreased and as the number of binned pixels is increased. This behavior is due to the fact that at slower acquisition rates, the maximum  frequency for technical noise that can be canceled by the subtraction of two consecutive frames is reduced.  This leads to a larger amount of classical excess technical noise after the subtraction, which can dominate over the spatial squeezing. Through the use of a simple model we are able to show that this is due to the temporal technical noise and not spatial inhomogeneities in the quantum efficiency of the EMCCD.

We believe that this comprehensive study of the limiting factors for measuring spatial squeezing in bright twin beams with an EMCCD camera will be helpful to study of the spatial quantum properties of bright twin beams in general and their applications to different quantum protocols \cite{RMP}. Moreover, the ability to obtain spatial quantum noise reduction with a macroscopic number of photons will find its way into applications such as enhanced image resolution in a single shot, which could be beneficial for biological and quantum-enhanced imaging.
\\
\\
This work was supported by the W.~M. Keck Foundation.

\end{document}